# Sn-loss effect in a Sn-implanted *a*-SiO$_2$ host-matrix after thermal annealing: A combined XPS, PL, and DFT study

D.A. Zatsepin[1,2], A.F. Zatsepin[2], D.W. Boukhvalov[3,4], E.Z. Kurmaev[1,2], N.V. Gavrilov[5]

[1] *M.N. Miheev Institute of Metal Physics, Russian Academy of Sciences, Ural Branch, 620990 Yekaterinburg, Russia*
[2] *Institute of Physics and Technology, Ural Federal University, 620002 Yekaterinburg, Russia*
[3] *Department of Chemistry, Hanyang University, 17 Haengdang-dong, Seongdong-gu, Seoul 133-791, Korea*
[4] *Theoretical Physics and Applied Mathematics Department, Ural Federal University, Mira Street 19, 620002 Yekaterinburg, Russia*
[5] *Institute of Electrophysics, Russian Academy of Sciences, Ural Branch, 62990 Yekaterinburg, Russia*

Amorphous *a*-SiO$_2$ host-matrices were implanted with Sn-ions with and without posterior thermal tempering at 900 °C for 1 hour in ambient air. X-ray photoelectron spectroscopy analysis (XPS core-levels, XPS valence band mapping), photoluminescence (PL) probing, and density functional calculations (DFT) were employed to enable a detailed electronic structure characterization of these samples. It was experimentally established that the process of Sn-embedding into the *a*-SiO$_2$ host occurs following two dissimilar trends: the Sn$^{4+}$→Si$^{4+}$ substitution in *a*-SiO$_2$:Sn (without tempering), and Sn-metal clustering as interstitials in *a*-SiO$_2$:Sn (900 °C, 1 hour). Both trends were modeled using calculated formation energies and partial densities of states (PDOS) as well as valence band (VB) simulations, which yielded evidence that substitutional defect generation occurs with the help of ion-implantation stimulated translocation of the host-atoms from their stoichiometric positions to the interstitial void. Experimental and theoretical data obtained coincide in terms of the reported Sn-loss effect in *a*-SiO$_2$:Sn (900 °C, 1 hour) due to thermally-induced electronic host-structure re-arraignment, which manifests as backward host-atoms translocation into stoichiometric positions and the posterior formation of Sn-metal clusters.

# 1. Introduction

Ion-implantation is a powerful technique for controllably managing embedded dopant concentration in a doped material [1-2], fabrication of non-stoichiometric defects that are typically employed in luminescing electronic devices [3], and selective modification of the surface and near surface layers of a host-material that dramatically re-builds the chemical properties of the inorganic material [4-5]. Although this technological method is relatively well-studied, developed and applied, the minority of unsolved problems still remains. The most important aspect during ion-implantation synthesis is ion-dopant allocation within the host-structure and the possibility of direct ion-replacement during the fabrication of substitutional impurities. In contrast with other conventional methods, such as molecular beam epitaxy or the sol-gel technique, when impurity incorporation into the host-structure occurs simultaneously on the stage of new matrix formation, the ion-implantation synthesis supposes "insertion" of the doping impurities into the unaffected host-structure of a crystal, amorphous solid or thin-film. Even if "inserted" atoms remain in the matrix as interstitials, they will impact the overall physical and chemical properties of a newly fabricated material. The situation appears to be even more complicated when multi-electron and multi-shell elements (such as Sn, Pb, Bi or similar) are implanted to yield wide-gape insulating matrices such as $a$-$SiO_2$. Such implants, being significantly valuable for concrete applicative employment, are impeding the precise characterization of the electronic structure of the final material through direct experimental techniques due to some issues, i.e. essential electron screening of relatively "deep" electronic levels with outer shells, electron-drift charging of the whole sample under study due to the loss of overall electro-neutrality caused by excitation sources in the spectrometer or another apparatus, etc. These reasons motivate an extension of experimental and theoretical studies regarding the various combinations of wide-gap amorphous hosts and impurities of multi-shell multi-electron type.

Technologically-modified $a$-SiO$_2$ matrices have been studied intensively in recent decades for various applications [6-12]. These studies have yielded new luminescing properties, single-electron effects, and a few optical nonlinearities, as well as establishing that structure imperfections (defects) that provoked nonlinear optical effects in the "defective" $a$-SiO$_2$ matrix (see Refs. [11-15]). The concentration of defects in $a$-SiO$_2$-hosts may be strongly decreased by post-annealing treatment that distorts the ordering type of embedded metal species as well as the microstructure of the final host-matrix. This is believed to be due to thermo-stimulated migration of oxygen and/or embedded metal-spice within the volume of a doped host [16-18] and, thus, the annealing mode has to be chosen with an essential accuracy in order to avoid or at least to minimize these side-effects.

Sn-ion implantation of silica glassy host-matrices is reported in Refs. [15, 19-21]. The greatest difference among the experimental data cited is that either only Sn-metal microcrystalline fractures of about 4-20 nm diameter or only uncontrolled oxygen self-diffusion in the host-matrix occur, depending on the implantation mode and post-implantation tempering conditions. As a result of self-diffusion, oxidation of embedded tin nanoparticles takes place in such a way that both Sn$^0$, Sn$^{2+}$ and some traces of Sn$^{4+}$ were simultaneously found. Additionally, no consensus has been reached regarding the ordering type of tin nanoclusters inside the host-matrix volume fabricated via the ion-implantation techniques [12]. Additionally the limited information regarding amorphous SiO$_2$ host-matrix reply to external treatment is usually reported, despite of the fact that various polymorphic structure re-arrangements are quite possible for silicon dioxide [22-24]. Thus, it is inconvenient to derive the preferable conditions of concrete embedding mechanisms for tin nanoparticles into $a$-SiO$_2$ in the form of only Sn-metallization (as a separate tin metal-phase), SnO-clusters (Sn$^{2+}$, on the cards and preferably as interstitials) or SnO$_2$-clusters (as expected Sn$^{4+}$→Si$^{4+}$ substitution or separate SnO$_2$-phase allocation) from these data. Considering the uncontrolled self-diffusion of oxygen under tempering, the embedding process for tin nanoparticles into silica glass seems to be essentially complicated and needs further study.

Therefore, Sn-incorporation into an $a$-SiO$_2$ matrix is, on the one hand, feasible for studying the allocation type of "replaced" host-atoms and thermal stability of doped samples under a low-concentration impurity mode, and on the other, quite valuable for the development of novel optical materials based on doped $a$-SiO$_2$. In this paper we present the results of an electronic structure study via X-ray photoelectron spectroscopy (XPS), photoluminescence (PL) probing, and density functional calculations (DFT) for $a$-SiO$_2$ host-matrices implanted with Sn-ions before and after thermal annealing. The mechanisms of Sn-incorporation into a silica glass host-matrix are suggested and discussed.

## 2. Experimental and calculation details

KU-type ("wet" sintering approach) quartz glass was used as a host-matrix for Sn-ion pulsed implantation (see for details Ref. [25]). The following ion-implantation stimulated synthesis was applied to the samples: implantation energy $E_{Sn}^+$ = 30 keV, fluence $\Phi$ = 5 × 10$^{16}$ cm$^{-2}$, ion-current density $J$ = 0.6 mA/cm$^2$, repetitively-pulsed mode with 0.4 ms pulse duration and a frequency of 25 Hz, post-implantation tempering at $T$ = 900 °C within 1 hour in the air. Both *as-is implanted* and *implanted and then thermally treated* samples have been analyzed and are reported here.

XPS survey (fast wide scan) and core-level analysis of the samples under study were made using a PHI XPS Versaprobe 5000 spectrometer (ULVAC-PHI, USA) with an Al $K\alpha$ X-Ray source (1486.6 eV) and energy resolution of $\Delta E \leq 0.5$ eV [25]. As the probing depth of the XPS is approximately 8 Å due to the inelastic mean free path (IMFP) of excited electrons [26], it is quite surface sensitive, being close to that of the total electron yield (TEY) absorption measurements. The XPS residual background was removed using the Tougaard method [27]. All spectra were calibrated using the reference energy value of the carbon core-level energy (C 1$s$) of 285.0 eV [28-29] under ASTM 2735 Measurements and Calibrating Rules [30]. Recall that the authenticity of XPS analysis in the case of elements like Sn, Pb, and Bi (multi-shell and multi-electron elements) is affected not only by the sensitivity factor of the concrete element under XPS study, but also by the XPS electron attenuation length due to multi-electron screening [31]. The situation becomes even more exaggerated if the analyzed multi-shell multi-electron element is embedded into a wide-gap insulating material because of the significant differential charging of the sample due to the loss of photoelectrons and the posterior unpredictable drift-discharge. This might be avoided by the short-time multipoint X-ray exposure mode for XPS spectra recording, which was applied with the following summation and averaging of the XPS data obtained [25]. Also the ASTM standards for the XPS measurements were strictly implemented and then additional photoluminescence (PL) probing was performed for additional validation. The PL-spectra were recorded in the 1.5 – 6.0 eV range with the help of an ARC Spectra Pro-308i monochromator and R6358P photomultiplier (HAMAMATSU, Japan). The helium-cooled PL-measurements were carried out at $T = 10$ K.

First-principle-based modelling using density functional theory (DFT) was performed to determine the formation energies of various configurations of structural defects induced by Sn-ion implantation of $a$-SiO$_2$ and a theoretical description of the electronic structure. These calculations were performed with the use of the SIESTA pseudopotential code [32], a technique that has recently been successful in similar studies regarding impurities in SiO$_2$ [25]. All calculations were made employing the Perdew–Burke–Ernzerhof variant of the generalized gradient approximation (GGA-PBE) [37] for the exchange-correlation potential in the spin-polarized mode. A full optimization of the atomic positions was carried out, during which the electronic ground state was consistently found using norm-conserving pseudopotentials for the cores and a double-$\xi$ plus polarization basis of localized orbitals for Si, Sn, and O. The forces and total energies were optimized with accuracies of 0.04 eV Å$^{-1}$ and 1.0 meV, respectively. The calculations of the formation energies ($E_{form}$) were performed by considering the supercell, both with and without given defects such as oxygen vacancies. We used the $Si_{24}O_{48}$ supercell for current modeling of tin impurity interactions with the quartz-like matrix. The calculations of the formation energies were performed via the application of a standard formula,

$$E_{form} = [E(Si_nO_{2n}) - zE(Si) - zE(O_2)/2 + yE(Sn) - E(Si_{n-x}Sn_yO_{2n-z})],$$

where $E(Si_nO_{2n})$ is the total energy of a supercell before modification; $E(Si_{n-x}Sn_yO_{2n-z})$ is the total energy of the host-system after the removal of $x$ atoms of Si, the formation of $z$ oxygen vacancies, and the insertion of $y$ Sn-ions (please note that $x$ and $z$ for some configurations could be equal to zero); $E(Si)$ and $E(Sn)$ are the energies of bulk Si and Sn in the ground state, respectively; and $E(O_2)$ is the total energy of a single oxygen gas molecule in the triplet state.

## 3. Results and Discussion

Figure 1 displays the XPS survey spectra recorded and calibrated under conditions reported in the Experimental and Calculation Details section. This figure clearly shows that the spectra of implanted and treated samples are exhibiting additional XPS peaks, identified as Sn 3*d* and Sn 3*p* states, as compared with *a*-SiO$_2$ external XPS standard Survey. No extra signals were found within the sensitivity range of the applied XPS spectrometer and XPS measurements mode, indicating that Sn-embedding into the target host was performed without additional contamination. The C 1*s* signal employed for calibration is relatively low, nevertheless allowing precise calibration [30] of the recorded spectra and recognition of the origin of the other XPS peaks. The Binding Energy (BE) region from approximately 20 eV to 25 eV in the XPS Survey of the $SiO_2$:Sn and $SiO_2$:Sn (900 °C, 1 hour) samples with high-probability relates to the overlapped Sn 4*d* – O 2*s* states (Fig. 1). This overlapping will be analyzed in detail within this paper using onward XPS valence band mapping and DFT-calculations.

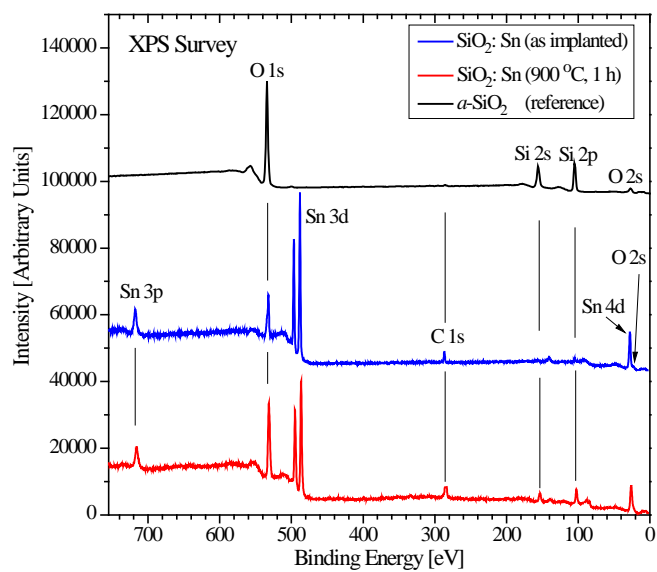

**Figure 1.** X-ray photoelectron (XPS) survey spectra for the *as-is* Sn-ion implanted *a*-SiO$_2$ host, *implanted and thermally treated* at 900 °C (1 hour), and *a*-SiO$_2$ external XPS standard.

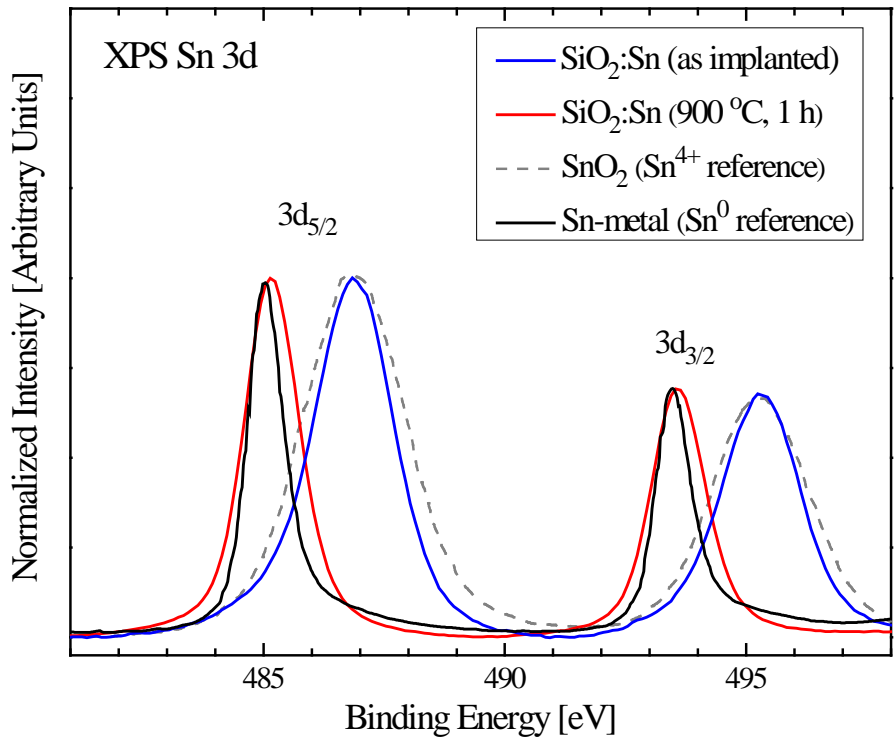

**Figure 2.** X-ray photoelectron (XPS) Sn 3*d* core-level spectra for the *as-is* Sn-ion implanted *a*-$SiO_2$ host, *implanted and thermally treated* at 900 °C (1 hour), and external XPS standards – $SnO_2$ ($Sn^{4+}$) and Sn-metal ($Sn^0$).

XPS Sn 3*d* core-level spectra for the *as-is* Sn-ion implanted *a*-SiO$_2$ host, *implanted and thermally treated* at 900 °C (1 hour), and the appropriate external XPS standards are shown in Fig. 2. The most interesting result is that the *as-is* Sn-ion implanted sample exhibit Sn 3*d* core-level spectrum with the same BEs of $3d_{5/2}$ – $3d_{3/2}$ peaks in the SnO$_2$ reference (as well as the spectrum shape) [28-29], indicating that the formal valency of tin after *as is implantation* is 4+. An employed thermal treatment, tempering the Sn-ion implanted sample at 900 °C within 1 hour, results in decreased Sn $3d_{5/2}$ – $3d_{3/2}$ BEs and transformation of the shape of the Sn 3*d* core-level spectrum so that it resembles that of Sn-metal (Sn$^0$). The specific behavior mentioned indicates that this mode of annealing simply destructs the tin–oxygen chemical bonding, previously fabricated via *as-is implantation*, and allocates the implanted Sn as metal-clusters. If our point of view is correct, we will find the appropriate transformations in the other XPS data for the samples under study and this will be reported and discussed.

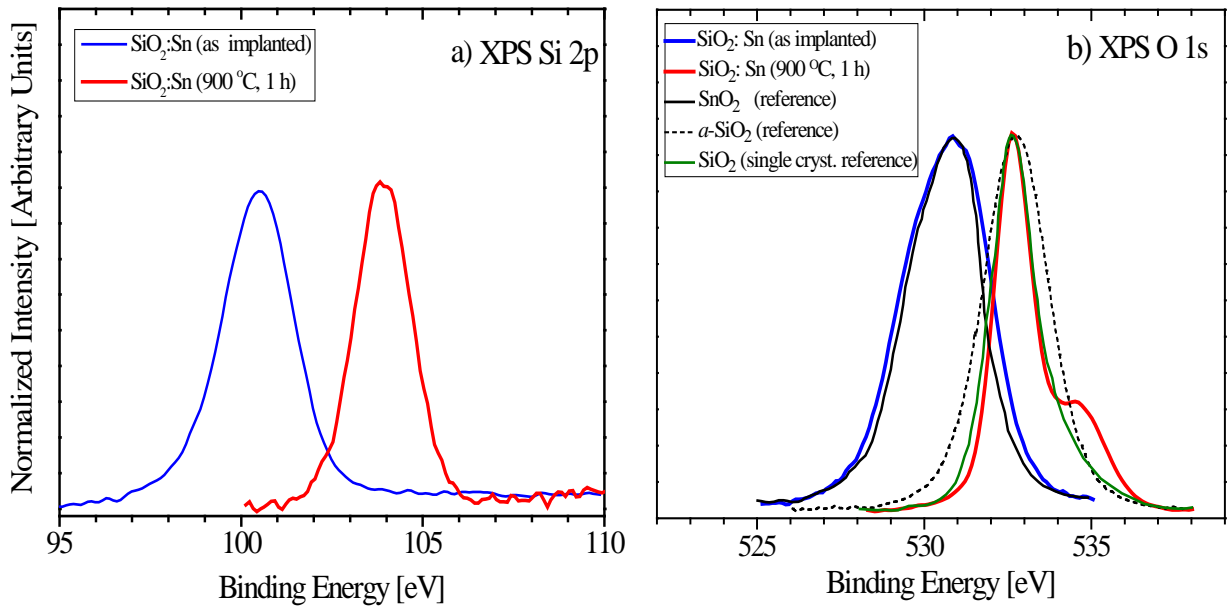

**Figure 3.** (a) X-ray Photoelectron (XPS) Si 2*p* core-level spectra for the *as-is* Sn-ion implanted *a*-SiO$_2$ host, *implanted and thermally treated* at 900 °C (1 hour); (b) X-ray Photoelectron (XPS) O 1*s* core-level spectra for *as-is* Sn-ion implanted *a*-SiO$_2$ host, *implanted and thermally* treated at 900 °C (1 hour), SnO$_2$, *a*-SiO$_2$ host and single crystal SiO$_2$ external XPS standards.

Figure 3(a) displays the XPS Si 2*p* core-level spectra of the samples under study. They are nearly the same in shape, but vary remarkably in BEs: 100.5 eV for the *as-is implanted a*-SiO$_2$ host and 103.8 eV for the *implanted and thermally treated* sample. According to the NIST Standard Reference Database [29], the BE of the *as-is* implanted *a*-SiO$_2$ host is close to that of decomposed *a*-SiO$_2$ matrices (Si 2*p* BE = 100.1 eV), denoted in this Database as the Si$_x$O$_y$ - Si system (silicon deficient matrix). At the same time, the Si 2*p* BE for the O$_2$/Si system affords 100.4 eV, where silicon is not completely oxidized — either it is located with oxygen surroundings initially without bonding with this oxygen, or it is located with the dangling oxygen bonds [34, 35]. This coincides with the obtained value of the binding energy for the *as is* Sn-ion implanted sample without thermal annealing (see Fig. 3a). As for the sample tempered at 900 °C, the BE for the Si 2*p* core-level is the same as that in SiO$_2$ [28]. The most interesting situation arises in the XPS O 1*s* core-levels (Fig. 3b); as can be seen in this figure, the O 1*s* core-levels for the *as-is* implanted *a*-SiO$_2$:Sn and SnO$_2$ (Sn$^{4+}$-reference) [36] are identical both in shape and BE location. If we also take into consideration the identity between the XPS Sn 3*d* core-level spectra for that sample and SnO$_2$, it seems that the embedded Sn directly replaces the Si-atoms within the glassy network of the *a*-SiO$_2$ host in the way of Sn$^{4+}$ → Si$^{4+}$, dangling the Si–O bonds and "shifting" the Si-atoms to the interstitial positions, thus allocating the silicon into a separate phase as in the O$_2$/Si system [34, 35]. This point of view is well supported by our XPS data for the Si 2*p* core-level presented in Fig. 3(a) and analyzed above.

As for 900 °C tempered *a*-SiO$_2$:Sn, again the notable shift in BEs indicates strong re-arrangement of the oxygen subnetwork in this sample (see Fig. 3b). It is clearly shown that 532.7 eV O 1*s* binding energy position coincides with that for the *a*-SiO$_2$ and SiO$_2$ single crystal, though our spectrum is significantly narrower in shape than O 1*s* in *a*-SiO$_2$. At the same time, a nearly perfect match occurs for the O 1*s* core-level of an SiO$_2$ single crystal, with the exception of the shoulder in our spectrum at nearly 534 eV. According to Refs. [37, 38], this shoulder is usually present in the XPS O 1$s$ spectra of $a$-SiO$_2$ matrices, which have been sintered employing so-called "wet" technology processing, and is related to the (OH)-group signal; this is not surprising, as we applied the same sintering approach to our initial $a$-SiO$_2$ hosts [25]. Thus, accumulating all the presented XPS data above, we can assume that annealing $a$-SiO$_2$:Sn samples at 900 $^o$C damages the fabricated SnO$_2$ clusters, allocating Sn-atoms in the form of metal clusters and re-arranging the Si–O bonding in such way that the amorphous host-matrix becomes more ordered when compared with the initial one. Nearly the same effect for "wet" $a$-SiO$_2$ matrices grown on Si-substrates but implanted with Ge-ions was reported earlier in Ref. [39]. These authors established that annealing of the Ge-implanted "wet" $a$-SiO$_2$ host results in damaging their samples and the occurrence of the Ge-loss effect, which leads to the oxygen dangling of Ge–O bonds and Ge allocation in the form of Ge$^0$ majority, and, to a lesser degree, GeO$_x$. Our XPS data show that we achieved the same loss-effect for Sn because of the annealing and applied "wet" technology of $a$-SiO$_2$ host synthesis.

X-ray photoelectron spectra of valence bands (VB) for the samples under study and the β-SnO$_2$ single crystal (Sn$^{4+}$-reference) are presented in Figs. 4 (a-b), where appropriate XPS VB external standards were taken from the internationally-accepted NIST XPS Standard Reference Database [29] and Refs. [40-42] for comparison. Onward XPS VB mapping was performed according to the references mentioned.

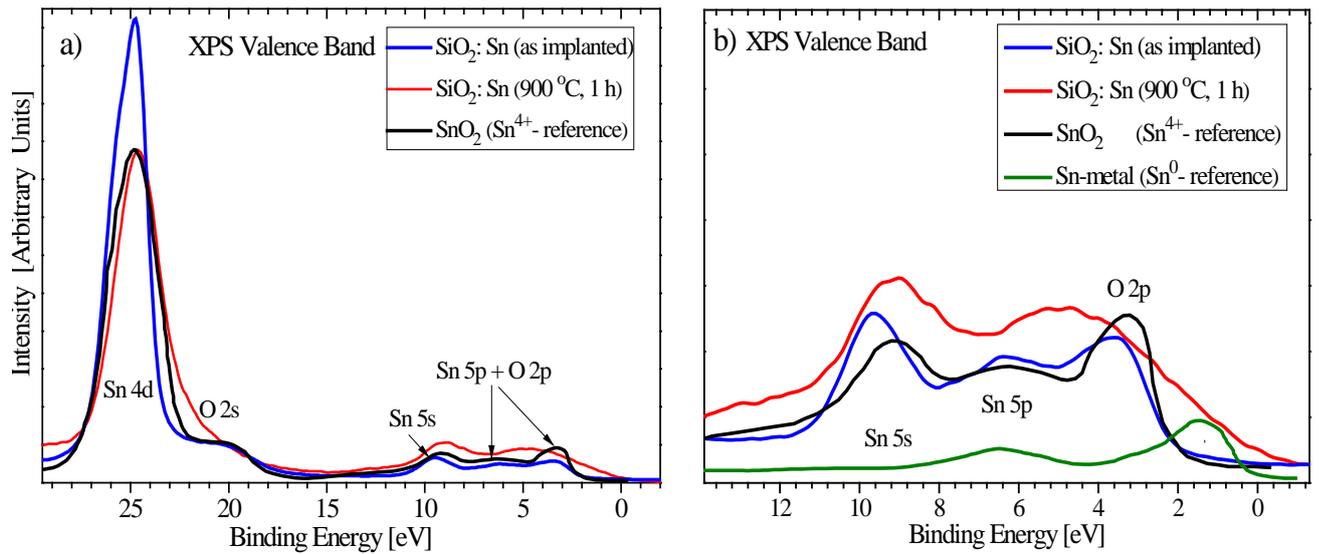

**Figure 4.** (a) X-ray photoelectron valence band (XPS VB) spectra of the *as-is* Sn-ion implanted *a*-SiO$_2$ host and the sample *implanted and thermally treated* at 900 °C (1 hour). The XPS valence band spectrum of the β-SnO$_2$ single crystal is shown as an external XPS standard [29]; (b) the zoomed XPS valence band area for the same samples. The Sn-metal XPS valence band spectrum is shown as an external XPS standard [44].

The strong intensive band located at ~ 25.2 eV belongs to Sn 4*d* states, whereas the low-intensity shoulder in the range of 19.8 – 22.3 eV relates to O 2*s* states [29, 42]. A comparably wide and low-intensity 10 eV band arose due to the majority of Sn 5*s* states because of only the band-tail and hence, very weak O 2*p* state contribution [43-44]. In both of these references, the area ~ 3 eV wide with the center located at ~ 6.2 eV is interpreted as a contribution of the hybridized Sn 5*p* – O 2*p* partial states to the valence band, and this point of view agrees with the DFT-calculations presented in these references. The interpretation mismatch between Refs. [39-40] arises while explaining the ~ 3.7 eV bands that are situated at the vicinity of $E_F$ – here, Köver et al. [41] assigned the origin of the XPS band mentioned above to the strong hybridization between O 2*p* – Sn 4*d* partial densities of states (PDOS). This conclusion contradicts the NIST XPS Standard Reference Database [29] and our XPS data, because it is known that Sn 4*d* and

O 2*s* electronic states are strongly dominant at deeper energy levels. The same point of view is presented in Ref. [46], where the ~ 3.7 eV band located at the vicinity of $E_F$ is explained as a contribution of O 2*p* – Sn 5*p* hybridized partial densities of states with a dominating contribution of O 2*p* PDOS. This interpretation is supported by XPS experimental data for $SnO_2$-based compounds [46], and seems to us more reasonable than the interpretation of Köver et al. [41]. The dominating O 2*p* state character of the ~ 3.7 eV band is well supported by our XPS data because this band is strongly sensitive to thermal annealing, which transforms its intensity and shape in such way that a vicinity tail at ~ 1.5 eV appears in the XPS VB spectrum for the *a*-$SiO_2$:Sn sample tempered at 900 °C (see Figs. 4 (a, b)). As discussed above, this might be a signature of oxygen migration due to annealing and the allocation of Sn as a metal spice in the volume of the implanted and tempered *a*-$SiO_2$ host-matrix. The dissimilarity in the spectral parameters of the ~ 3.7 eV XPS band among the reference β-$SnO_2$ and *as is implanted a*-$SiO_2$:Sn is due to the use of the single crystal of β-$SnO_2$ as an XPS reference standard, while the samples under study are of amorphous origin from the beginning. Nearly the same difference was found in the XPS O 1*s* core-level spectra between the *a*-$SiO_2$:Sn, *a*-$SiO_2$:Sn (900 °C, 1 hour), and $SiO_2$ single-crystal external XPS standard (see Fig. 3b). Thus, the performed XPS valence band mapping analysis on the whole coincides with the XPS core-level analysis presented above. The band-tail intensity at approximately 12.2 eV, clearly seen in the VB spectrum of *a*-$SiO_2$:Sn (900 °C, 1 hour), is absent in the reference β-$SnO_2$ and thus, cannot be identified on the basis of Refs. [41, 42], so the origin of this band will be clarified on the basis of electronic structure calculations.

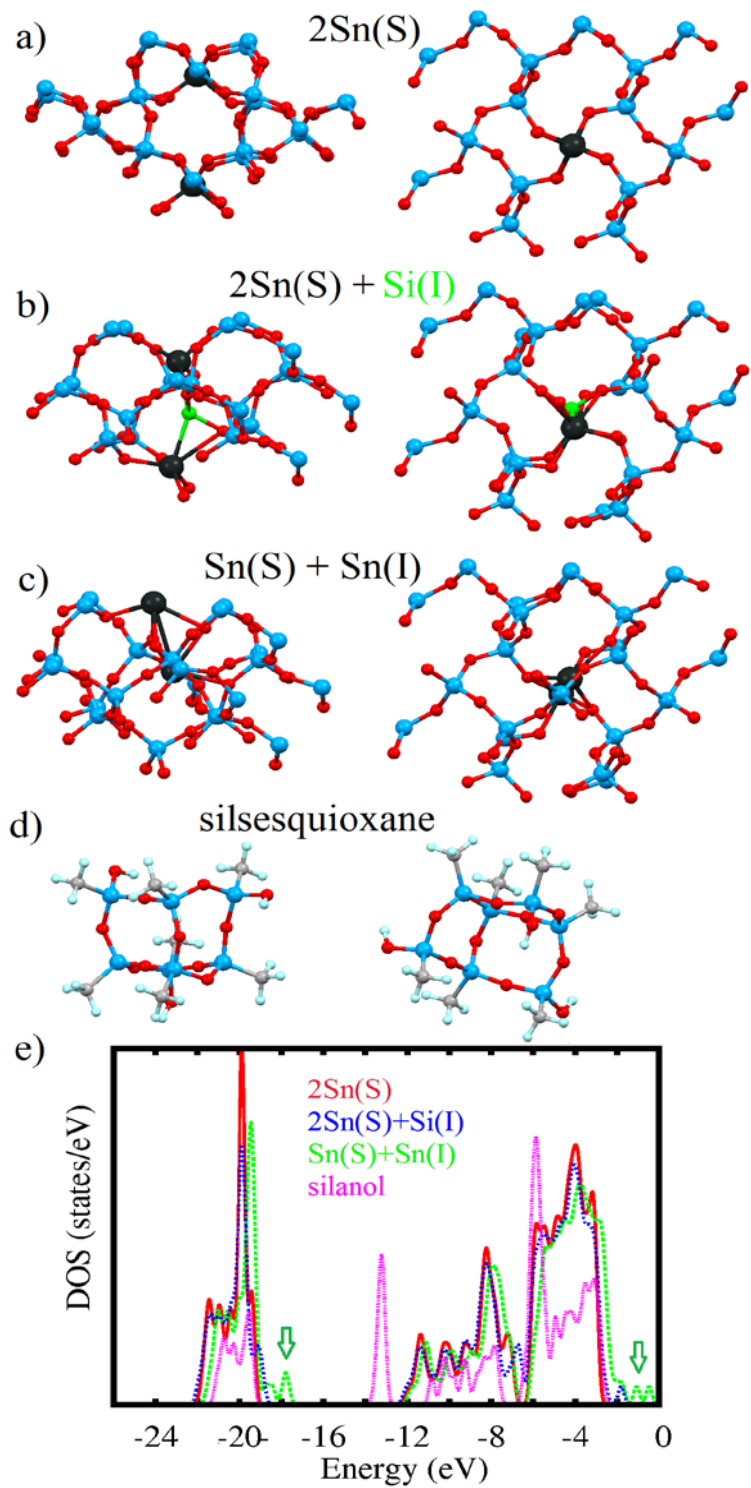

**Figure 5.** Optimized atomic structure for various configurations of defects in the SiO$_2$ matrix (a-c), silanol (d) and the total densities of states per atom corresponding to presented atomic structures.

Based on our previous experience in the field of modeling various impurity types in $SiO_2$ [25] and other oxide matrices [44], we have studied the formation energetics for the most probable variants of substitutional (*S*) and interstitial (*I*) impurities, as well as their combinations (see Figs. 5 (a, b)) with oxygen vacancies (vO). In contrast to our previous calculations, we also take into account additional opportunities for $Sn^{4+} \rightarrow Si^{4+}$ substitution with the simultaneous appearance of an Si-atom in the interstitial void (Fig. 5 (c)). Our calculations of the energetics for defects formation (Tab. I) demonstrate that in the absence of any additional defects, the most energetically favorable is the configuration assembled from the pair of substitutional and interstitial Sn-impurities (Fig. 5 (b)). The presence of usually unavoidable oxygen vacancies makes the formation energies for all types of defects almost the same. The presence of an additional Si-atom in the interstitial void near the Sn-impurities (Fig. 5 (c)) dramatically changes the energetics of "insertion" and makes combinations like *x*Sn(S)+I(Si) more likely (see Tab. I). The theoretical results obtained support the background experimental data discussed above as evidence for the presence of Si-atoms not only in the stoichiometric positions in $SiO_2$:Sn.

**Table I.** Formation energies (in eV per defect) for various configurations of impurity atoms. The most probable configurations are indicated in bold.

| Configuration | only Sn-defect | +$V_O$ | +I(Si) |
|---|---|---|---|
| 1*S* | 5.38 | 4.26 | **2.89** |
| 2*S* | 5.35 (Fig. 6a) | 4.62 | **2.88** (Fig. 6c) |
| *I* | 5.67 | 5.21 | 3.72 |
| *S*+*I* | **2.96** (Fig. 6b) | 4.62 | **2.75** |
| 2*S*+*I* | 4.44 | 4.15 | 3.04 |

The calculations for the $Sn^{4+} \rightarrow Si^{4+}$ substitution of two Si-atoms, it is clear that this also does not alter the visible electronic structure of $SiO_2$, which "remains" quartz like (Fig. 5 (e)). Thus we can conclude that coordination of for-valent ions in play an important role for them electronic structure and without changes of local environment changes in electronic structure will be minor. An additional insertion of Si-atoms into the interstitial void (2*S*(Sn)+*I*(Si), Fig. 5 (c)) does not provide any valuable contribution to the electronic structure of $SnO_2$ because minor changes disappear through thermal smearing. Thus, we can conclude that as in experimental valence band mapping (Fig. 4b), the theoretical calculations also demonstrate the absence of influence on the electronic structure of valence bands due to the movement from stoichiometric positions to interstitial void Si-atoms by means of Sn-impurity. In contrast to the case in which Sn-impurities occupy only stoichiometric positions, the combination of these impurities in substitutional and interstitial positions (Fig. 5b) provides essential re-arrangements in electronic structure, as in the broadening of the O 2*s* peak and the appearance of metal-like states at approximately -1 eV (see arrows in Fig. 5e). This changes in electronic structure can be noted as first step of transformation of quartz-like electronic structure to rutile-like by increasing of coordination number of Sn-impurities from 4 to larger values. Thus, from the results of theoretical modeling, we might conclude that tempering induces an exchange of atoms in stoichiometric and non-stoichiometric positions and the atoms of the host-matrix return to their initial sites, shifting Sn-impurity atoms to the interstitial void. Note that in the reported synthesis of $SiO_2$:Sn (900 $^o$C, 1 hour), the annealing takes place in ambient air, which coincides with the treatment of oxygen vacancies. The latter is also in agreement with the results of DFT-modeling, when the *S+I* configuration is significantly energetically favorable only when the oxygen vacancies are completely treated. Unfortunately, the results of our calculations do not explain all of the transformations of electronic structure during annealing, and particularly the peaks at approximately 6.2 and 12.2 eV, which cannot be caused neither by appropriate DOS contribution from $SiO_2$ nor by pure Sn-metal (see Fig. 4b). Thus, we took into account the amorphous origin of our samples, allowing us to simulate the formation of silanol-containing structures – *silsesquioxane* - which might be considered as a feasible model for the abovementioned case of matter. The results of our calculations demonstrate that the VB peaks for the annealed samples might be related to silanol-containing structures that have been fabricated via the applied thermal treatment. In terms of this model approach, the wide and low-intensity tail-band at 12.2 eV (in

Fig. 4b) in experimental VB for $a$-SiO$_2$:Sn (900 °C, 1 hour) is possibly the tail from a silsesquioxane-like group, partially containing OH-groups, because the KU-SiO$_2$ matrix was used for sintering [25]. Hence, we assume that *silsesquioxane* DOS contribution also occurs in the 6.2 eV VB band of the aforementioned sample.

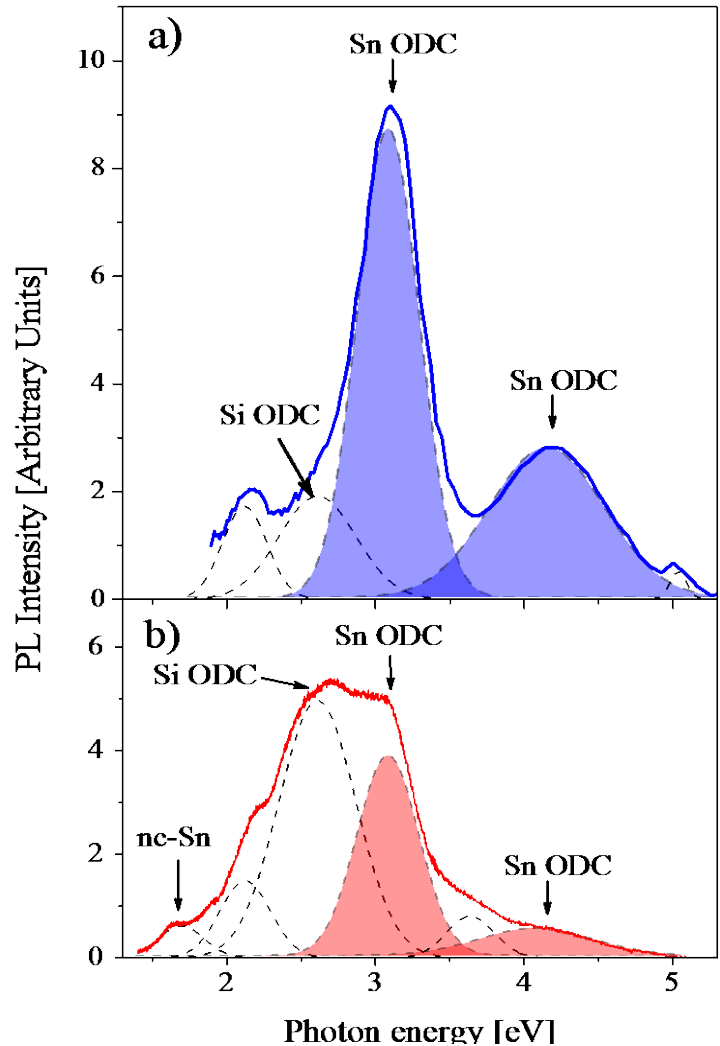

**Figure 6.** Photoluminescence (PL) spectra of Sn-ion implanted KU-$a$-SiO$_2$ glasses measured at 10 K (a) before and (b) after thermal annealing.

An implementation of the Sn-embedding effect into the glassy network of $a$-SiO$_2$ (KU-type) as Sn$^{4+}$ → Si$^{4+}$ substitution, the posterior structural re-arrangement of tin particles due to the tempering conditions, as well as the possibility of fabrication of isolated (Sn)$_n$ nanoclusters, are all confirmed by the PL-probing data shown in Fig. 6. These spectra were deconvoluted using Gaussian components in order to clearly separate the contribution of luminescence bands with dissimilar radiative relaxation mechanisms. In this figure, there are two high-intensity luminescence bands located at 3.1 eV and 4.1 eV that exhibit the spectral parameters presented in Table II.

**Table II.** Spectral parameters of PL-bands for *a*-SiO$_2$:Sn

| PL-center | Transition | As is Sn-implanted | | After annealing at 900 °C | |
|---|---|---|---|---|---|
| | | h$\nu_m$ (eV) | FWHM (eV) | h$\nu_m$ (eV) | FWHM (eV) |
| Sn ODC | T$_1 \rightarrow$ S$_0$ | 3.1 | 0.49 | 3.1 | 0.49 |
| | S$_1 \rightarrow$ S$_0$ | 4.17 | 0.9 | 4.1 | 0.9 |
| Si ODC | T$_1 \rightarrow$ S$_0$ | 2.62 | 0.6 | 2.62 | 0.6 |

These parameters correspond to luminescing transitions in tin-containing oxygen deficient centers (Sn ODC) that are of defective origin [14]. The 2.65 eV PL-band is linked to similar isoelectronic luminescence centers such as the silicon-type Si ODC, whereas the relatively weak-intensity 1.6 eV PL-band appearing after thermal treatment at 900 °C arises due to the luminescence of (Sn)$_n$ clusters [14, 42]. As shown previously in Ref. [43], the most effective photoluminescence excitation of metallic tin nanoparticles is achieved through non-radiative energy transfer by excitons of an *a*-SiO$_2$ matrix.

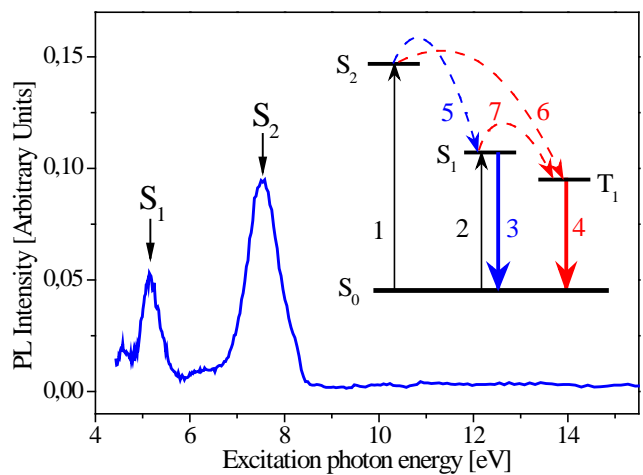

**Figure 7.** Photoluminescence spectrum of the excitation of a singlet 4.1 eV PL-band and the scheme of electronic transitions for Sn ODC in *a*-SiO$_2$:Sn without thermal annealing. S$_1$, S$_2$ and T$_1$ denote the PL-excited states of the Sn ODC defect.

A diagram of optical transitions in Sn ODC as well as appropriate experimental PL-spectrum is shown in Fig. 7. The 5.15 eV and 7.6 eV PL-bands correspond to transitions 1 and 2, respectively, which are linked to the excitation of defective Sn ODC to the primary ($S_1$) and secondary ($S_2$) singlet states. Radiative transitions 3 and 4 in Fig. 6 characterize the PL-bands of singlet-singlet (4.1 eV) and triplet-singlet (3.1 eV) luminescence (see Fig. 6). Finally, transitions 5 and 6 illustrate the occupation of singlet $S_1$ and triplet $T_1$ states under the excitation of Sn ODC luminescence at the 7.6 PL-band. The transition 7, denoting the triplet $T_1$ state, is thermally activated and displays the occupation of this state under excitation of the 5.15 eV PL-band $S_1$ for Sn ODC.

As shown in the PL-spectra of the Sn ODC presented in Fig. 6, the triplet 3.1 eV luminescence always has higher intensity than that of the singlet 4.1 eV PL. This indicates that while exciting the $S_2$ state, the $S_2 \rightarrow T_1$ transitions are more probable than the $S_2 \rightarrow S_1$ transitions. This peculiarity is primarily realized in the presence of silicon ODC-defects, although for isoelectronic Si ODC at low temperatures ($T = 10$ K), it is possible to register only the triplet 2.65 eV PL-band (Fig. 6), which highlights the existence of an energy barrier among the excited $S_2$ and $S_1$ states. Finally, from our luminescence data, it follows that during Sn-implantation, the predominant embedding of tin atoms into silica atom positions occurs in the structure of the *a*-$SiO_2$ matrix. This process takes place directly with the formation of oxygen deficient clusters of ODC type, including a relatively small quantity of similar ODCs of silicon nature. This stage is indicated by the appearance of intensive Sn ODC luminescence in the samples without thermal annealing (*as-is* implanted). A remarkable decrease (more than 3 times) in Sn ODC luminescence after thermal annealing denotes the re-arrangement process of tin particles, namely, the backward structural transition from silicon positions to interstitials with posterior clustering.

On the whole, one might suppose that Sn-ions embedding in the following thermal treatment at 900 °C have a significant influence on the structure of amorphous glassy $SiO_2$ matrices with the formation of luminescing nanosized $(Sn)_n$ metallic clusters, nanoparticles in the form of $SnO_2$, ballistic damaged parts of glassy networks like shifted Si, and O atoms due to Sn-implantation (recall that an implantation process usually consists of two stages – ballistic interactions and implantation-stimulated chemical synthesis [4]). While tempering the implanted $a$-$SiO_2$:Sn sample, the processes of allocation of tin-containing nanoparticles and thermal treatment of luminescing defects occur simultaneously. We note that the mechanisms of Sn-implant behavior in a glassy $SiO_2$ host-matrix reported in the current paper well agree with that reported in Ref. [46] for Si/$SiO_2$:Sn thin-film heterostructures.

**Conclusions**

The combined XPS, PL, and DFT characterization of the electronic structure of a Sn-implanted glassy $SiO_2$ host-matrix before and after thermal tempering (900 $^o$C, 1 hour) was performed. The employed approach allows clear detection and separation of the two dissimilar mechanisms of Sn-embedding into the $SiO_2$ host-matrix: the $Sn^{4+} \rightarrow Si^{4+}$ substitution by moving the host-atoms from stoichiometric positions into the intestinal void for samples without tempering, and the Sn-loss effect for the thermally-treated host-matrices. The latter is represented by partial backward translocation of host-atoms into their stoichiometric positions simultaneously with Sn-metal clustering as interstitials in the amorphous three-dimensional glassy network of the $SiO_2$-host. The key-point of employed thermal treatment for the essential re-arrangement of the electronic structure in the samples was observed in a study of their luminescence features.

**Acknowledgements**

The preparation of *a*-$SiO_2$ samples, ion-implantation treatment, and photoluminescence measurements were supported by the Russian Foundation for Basic Research (Projects RFBR Nos. 13-08-00568 and 13-02-91333), the Act 211 of the Government of the Russian Federation (Contract № 02.A03.21.0006), and the Government Assignment of the Russian Ministry of Education and Science (3.1016.2014/K). The XPS measurements were supported by the Russian Science Foundation (Project No. 14-22-00004).